\documentstyle[pre,aps,psfig,epsfig,preprint]{revtex}

\begin{document}

\title{
LOCALIZATION AND EQUIPARTITION OF ENERGY IN THE $\beta$-FPU CHAIN :
CHAOTIC BREATHERS
.}
\author{
Thierry Cretegny$^{\diamond,\S}$,
Thierry Dauxois$^{\diamond,\S}$, 
Stefano Ruffo$^{\diamond}$\thanks{INFN and INFM, Firenze (Italy)}
and Alessandro Torcini$^{\diamond}$\thanks{INFM, Firenze (Italy)}}
\address{$^{\diamond}$ Dipartimento di Energetica "S. Stecco", 
Universit\`a di Firenze,
via S. Marta, 3 , I-50139 Firenze, Italy}
\address{$^{\S}$ Laboratoire de Physique, URA-CNRS 1325,
ENS Lyon, 46 All\'{e}e d'Italie, 69364 Lyon C\'{e}dex 07, France}
\date{\today}
\maketitle
\begin{abstract}

The evolution towards equipartition in the $\beta$-FPU chain is
studied considering as initial condition the highest frequency
mode. Above an analytically derived energy threshold, this
zone-boundary mode is shown to be modulationally unstable and to give
rise to a striking localization process. The spontaneously created
excitations have strong similarity with moving exact breathers
solutions.  But they have a finite lifetime and their dynamics is
chaotic.  These chaotic breathers are able to collect very efficiently
the energy in the chain. Therefore their size grows in time and they
can transport a very large quantity of energy.  These features can be
explained analyzing the dynamics of perturbed exact breathers of the
FPU chain. In particular, a close connection between the Lyapunov
spectrum of the chaotic breathers and the Floquet spectrum of the
exact ones has been found. The emergence of chaotic breathers is
convincingly explained by the absorption of high frequency phonons
whereas a breather's metastability is for the first time identified.
The lifetime of the chaotic breather is related to the time necessary
for the system to reach equipartition.  The equipartition time turns
out to be dependent on the system energy density $\varepsilon$
only. Moreover, such time diverges as $\varepsilon^{-2}$ in the limit
$\varepsilon \to 0$ and vanishes as $\varepsilon^{-1/4}$ for
$\varepsilon \to \infty$.

\end{abstract}
\vskip 3truecm

\section{INTRODUCTION}

In 1955, in one of the first but well known numerical simulations,
Fermi, Pasta and Ulam (FPU)~\cite{FPU} have observed the {\it absence}
of thermalization in a nonlinear lattice in which the energy was
initially fed into the lowest frequency mode.  Even if a lot of
progress have been made~\cite{ford} in the last thirty years on the
study of the evolution towards energy equipartition among linear
normal modes, several points are far from being clarified. For
historical reasons, the evolution towards equipartition has been
usually analyzed considering an initial state where all the energy of
the system was concentrated in a small packet of modes centered around
some low frequency~\cite{ford,ll}.  Only a few studies have been devoted
to the evolution from an initial condition where all the energy was
fed in the highest frequency mode \cite{maledetirusi,sandusky}.  From
these analyses it turned out that this different initial condition
leads to a completely new dynamical behavior in the transient time
preceding the final equipartition.  In particular, the main finding
was the appearance of a sharp localized mode during the transient
\cite{maledetirusi}. 
Moreover, the highest frequency mode turns out to be a linearly
unstable periodic solution; thus, starting the orbit from this initial
point in phase-space put the system on a hyperbolic point embedded
into a chaotic layer. On the other hand, when low frequency modes
are initially excited, instability arises because of the presence of 
thin chaotic layers near elliptic points~\cite{Chirikov}. Therefore,
we expect that the time-scales for the relaxation to the equipartition
state and the physical picture of the evolution towards the final 
state should be strongly affected.

In this paper, we present a detailed numerical and theoretical
study of the dynamics of a $\beta$-FPU model in the transient
preceding equipartition, when only the highest frequency mode is
initially excited.  Employing the quite recent concept of breather
excitations\cite{mackayaubry}, we are able to give a more detailed
explanation of some of the behaviors observed in
Ref.~\cite{maledetirusi}.  In particular, we show that the
existence of such localized modes during the transient is strongly
connected to exact breather modes for the $\beta$-FPU model
\cite{flachwillis}.  An important peculiarity of these excitations is
that (contrary to exact breathers) they have a chaotic evolution in
time, therefore we have termed them ``chaotic breathers'' (CBs).  The
fact that localized oscillating excitations (that can be identified as
CBs) show up spontaneously~\cite{prllocali,daumont} and persist in
numerical simulations has suggested that they play an
important role in the dynamics of Hamiltonian anharmonic systems.
However, an important point that should be clarified is why these
objects emerge so easily in Hamiltonian systems.  Since Hamiltonian
dynamics is reversible, large packets should break up into smaller
ones at the same rate that small ones merge into larger one; while in
the formation of a CB the latter process seems to be favored.  It is
one of the purposes of this paper to give some clarification about
these points and in particular to emphasize the importance of these
localized excitations for the transition towards energy equipartition.

Another important aspect that we discuss is the existence 
of scaling laws for the indicators characterizing the
approach to the equipartition state. We notice that quite
general scaling laws indeed exist in the thermodynamic
limit. Moreover, the relevant quantity for the equipartition time
turns out to be the energy density.

We have organized the paper in the following way. The results of
numerical simulations concerning the appearance of CBs are presented
in Sec.~\ref{localization}.  The consequences of the existence of
CBs for the transition to equipartition are presented in
Sec.~\ref{transition} together with the observed scaling laws.  The
relation between CBs and exact breathers is discussed in detail in
Sec.~\ref{exact}. Finally, Sec.~\ref{Formationdestruction} will deal
with the mechanisms of creation and destruction of breathers.  Some
final remarks and conclusions are reported in Sec.~\ref{conclusions}.

\section{MODULATIONAL INSTABILITY AND ENERGY LOCALIZATION}
\label{localization}

Denoting by $u_n(t)$ the position of the $n$th atom ($ n \in [1,N]$),
the equations of motion of the $\beta$-FPU chain read 
\begin{eqnarray}
\ddot{u}_n = u_{n+1}+ u_{n-1}- 2u_n +\beta\ \left[(u_{n+1}-u_n)^3
-(u_n - u_{n-1})^3\right] \label{sub} 
\end{eqnarray}
where $\beta=0.1$. The parameter $\beta$ can be absorbed with an
appropriate rescaling of $u_n$, but we keep it not only for
historical reasons, but also in order to make easier comparisons with
results reported in previous papers.  We have chosen periodic boundary
conditions which allow the propagation of waves in the lattice. As
recently shown in \cite{PRK}, the study of the evolution of traveling
waves with wavelength of the order of the system size (i.e. $ \simeq
N$) can give very useful information.  In particular, the average
lifetime of such waves is strictly related to the time necessary to reach
equipartition. Here again, the propagation of waves and localized
structures will play a fundamental role in the evolution of the
system.

As usual, in numerical simulations one does not study the real
Hamiltonian system, but a discrete time version that approximates the
time continuous dynamics. It is therefore essential for long time
simulations to use an appropriate symplectic integration scheme in
order to preserve as far as possible the Hamiltonian structure of the
problem. We adopt the 6th-order Yoshida's algorithm~\cite{yoshida}
with a time step $dt = 0.01$; this choice allows us to obtain an 
energy
conservation with an accuracy $\Delta E \simeq 10^{-11}$ (that
corresponds to a relative accuracy $\Delta E / E$ ranging from
$10^{-10}$ to $10^{-12}$).

We follow the approach proposed by Fermi-Pasta-Ulam in
Ref.~\onlinecite{FPU} where they look at the stability of one normal
mode of the harmonic part, but contrary to them we have performed
simulations adopting as initial condition the highest frequency mode.
The highest frequency mode (the so called $\pi$-mode) corresponds to
the following zig-zag pattern for $u_n$
\begin{equation}
u_n=(-1)^n\ a\quad \hbox{and}\quad \dot u_n=0
\label{pimode}
\end{equation}
where $a$ is its amplitude.  Since most of the normal modes of the
harmonic part of the Hamiltonian are no longer solutions of the full
Hamiltonian, energy initially fed into one single mode will be shared
on later times among other modes.  However, this is not true for three
particular modes that are exact solutions of the FPU
lattice~\cite{PR}. As the $\pi$-mode is one of these solutions, in
order to destabilize such initial state a small amount of noise (of
order $10^{-14}$) has been added on the velocities.

As already shown in Refs.~\cite{sandusky,prl} the $\pi$-mode turns out
to be modulational unstable above a critical energy $E_c$, that can be
analytically derived~\cite{prl}

\begin{equation}
E_c = {{2 N} \over {9 \beta}} 
\sin^2\left({{\pi}\over{N}} \right)
{{7 \cos^2({{\pi}\over N})-1}
\over
{(3 \cos^2({{\pi}\over N})-1)^2}} \quad .
\label{ec}
\end{equation}
As for this mode, $E=N(2a^2+4\beta a^4)$,
we easily see that if 
\begin{equation}
a>a_c=\sin(\pi/N)/\sqrt{\beta(9\cos^2(\pi/N)-3)}
\label{ac}
\end{equation}
the $\pi$-mode  will be destroyed by modulational instability.

In Fig.~\ref{greyscales}(a) a generic evolution of the above
initial state for $ a > a_c$ is reported. The grey scale refers 
to the energy residing on site~$n$,
\begin{equation}
E_n = {1\over 2} \dot{u}_n^2 + {1\over 2}V(u_{n+1}-u_n) +
{1\over 2} V(u_n-u_{n-1})\quad,
\end{equation}
where the substrate potential is $V(x) = {1\over 2}x^2 + {\beta\over
4} x^4$.  Figs.~\ref{greyscales}(b), \ref{greyscales}(c) and
\ref{greyscales}(d) refer to three successive snapshots of the local
energy $E_n$ along the chain. After a very short delay, a slight
modulation of the energy in the system appears (see
Fig.~\ref{greyscales}(b)) and the $\pi$-mode is
destabilized. Later, as attested by Fig.~\ref{greyscales}(a), only a
few localized energy packets emerge from this temporary state; they
correspond to oscillating localized waves and are usually called
breathers or intrinsic localized modes.  At this stage, as
inelastic collisions of breathers have a systematic tendency to favor
the growth of big breathers at the expense of small
ones\cite{prllocali,BangPeyrard}, the number of localized objects
decreases and only one very large amplitude breather-like excitation
survives (see Fig.~\ref{greyscales}(c)): this is the excitation we
have termed CB.  The CB moves along the lattice with a perturbed
ballistic motion: sometimes the CB is even stopped or
reflected. During its motion the CB collects energy from the visited
sites on the chain, and its amplitude increases. It is important to
note that the CB will never be at rest and that it propagates in
general with almost the same speed (in modulus). 
Finally, after a very long time
and through a mechanism we will consider later, the CB decays and the
system reaches the equipartition of energy, as illustrated in
Fig.~\ref{greyscales}(d).

The main important aspect that arises from the above reported 
picture is that the dynamics of the model seems to favour
the emergence of a well localized state (CB) during the transient
preceding equipartition \cite{maledetirusi}.
In order to give a more quantitative characterization of the
energy localization, we introduce the following quantity 
\begin{equation}
C_0(t)=N {\displaystyle \sum_{i=1}^N E_i^2\over 
\left(\displaystyle \sum_{i=1}^N E_i  \right)^2 }\quad;
\label{clocal}
\end{equation}

As it can be easily seen $C_0$ is of order one if $E_i = E/N$ at
each site of the chain and of order $N$ if the energy is localized on
only one site. In Fig.~\ref{coherence} $C_0$ is reported as a function
of time for an initial condition (\ref{pimode}) with $a >
a_c$. Initially $C_0$ grows in time, indicating that the energy,
evenly distributed on the lattice at $t=0$, localizes over
a few sites.  This localized state survives for some time,
at later times $C_0$  begins to decrease and finally it reaches 
an
asymptotic value $ \bar C_0$ that is associated with a total
disappearance of the CB.  At this stage a state with a flat
distribution of energy in Fourier space is attained, i.e. a state
where equipartition of energy is fulfilled.  Already at this point, it
is important to remark that for $N$ greater than 512, all the curves
$C_0(t)$ are almost coincident at any time. We discuss this point
in Sec.~\ref{Formationdestruction}.

The asymptotic value $\bar C_0$ can be easily obtained.
In the limit $t \to \infty$, the energy per site has a mean value
$\varepsilon=E/N$, but with some site dependent fluctuations, 
therefore $\bar C_0={\langle E_i^2\rangle/\langle E_i\rangle^2}$. 
$\bar C_0$ can be theoretically estimated
within a canonical ensemble picture, 
that allows us to derive an expression 
for $\langle E_i^2\rangle$ (more details are reported in Appendix A).
The actual $\bar C_0$-value depends on energy density
$\varepsilon$ and lies of course between the two limiting 
values: 
corresponding to the pure harmonic case $\bar C_0= 7/4$ and the 
pure quartic case $\bar C_0= 19/9$
(notice that both these values are independent of $\varepsilon$). 
As derived in Appendix A, for
$\varepsilon=1.44$, we obtain $\bar C_0\simeq 1.795$, 
in perfect agreement with the numerical results as shown in 
Fig.~\ref{coherence}.

\section{TRANSITION TO THE ENERGY-EQUIPARTITION STATE}
\label{transition}

In the previous section, we have shown that the energy-equipartition
state is preceded by a transient characterized by the emergence
of a localized state. The main properties of this state and its
relationship with exact breathers is studied in the next section.
Here we are mainly concerned with the identification of some
general aspects of the transient. In particular, scaling laws
for two indicators measuring the relaxation to thermodynamical
equilibrium are derived.

As a first indicator we consider a parameter in the
Fourier space $S(t)$ that gives a quantitative estimation
of the energy transfer among the different normal modes \cite{LPRSV}.
The energy associated with the Fourier mode $q= 2\pi k/N$, 
with $k \in\{1\ldots N/2\}$, is in the harmonic approximation
$\phi_q={1\over 2}\left(|V_q|^2+\omega(q)^2|X_q|^2\right)$,
where $\omega(q)=2\sin (q/2)$ represents the linear dispersion 
relation
and $V_q$ and $X_q$ are the Fourier transforms of the velocities and
the positions, respectively. 
Initially, all the energy is put in the $\pi$-mode, but as soon as the
modulational instability develops also the nearest modes 
$q = \pi - \delta q$ acquire a non-zero amplitude $\phi_q$, 
these numerical results confirm a previous analytical derivation
of the most unstable modes~\cite{prl}.
After this initial stage, energy transfer from the highest modes 
to the lowest ones continues
and the shape of the spectrum as a function of the wavevector is well
described by an exponential
\begin{equation}
\phi_q\sim\exp\left[-S(t) \left(\pi-q\right)\right]
\end{equation}
where, at any time $t$, $S(t)$ is the slope of the linear fit  in
the Lin-Log scale. 

When the equipartition of energy is reached, the
spectrum is no longer exponential (in other words the slope $S$
vanishes). Therefore, the slope $S$ is a excellent tool to
follow the transition towards equipartition. The numerical
results that we have obtained yield the interesting
conclusion that the equipartition time is a function of the
energy density $\varepsilon$ only, as attested  by 
Fig.~\ref{slope}(a), where
all evolutions of $S(t)$ for various lengths are almost
indistinguishable except for the smallest chain $N=32$ where
finite size effects do appear~\cite{alter}. 

Let us remark that in the original FPU problem, with
excitations of long-wavelength modes, the distribution was
well approximated by
$\phi_q\sim\exp\left[-S(t)\ q\right]$.
In that context, introducing the analytic continuation of
the continuum field $u(x,t)$ to the complex plane~\cite{LPRSV}, the
slope was directly related to the imaginary part of
the nearest singularity to the real axis. Work along this
line would probably lead to a similar conclusion in our case.

The transition to equipartition can be investigated even more 
precisely
considering the following indicator~\cite{Livi85}
\begin{equation}
\eta(t)={2\over N}\ e^{-\displaystyle\sum_{q} \ p_q(t)\ \ln p_q(t)},
\label{eta}
\end{equation}
where $p_q(t)$ is the probability to have an energy $\phi_q$
associated to the mode $q$ at time $t$; i.e.
\begin{equation}
p_q(t)={\phi_q\over\displaystyle \sum_{q}\phi_q}
\quad .
\end{equation}
Indeed, $\eta(0)=0$ whereas $\max(\eta)=1$ is reached~\cite{coreceta}
in the energy equipartition state where all $p_q=2/N$ (for a detailed
derivation of the equipartition value $\bar \eta$ see Appendix B). 
Thus energy
sharing among normal modes will be detected by an increase of 
$\eta(t)$,
which can be considered as the percentage of modes with significant
energy. It is exactly what we obtain in Fig.~\ref{slope}(b) for
various chain lengths. The main conclusion that we can draw from
this result is that again the equipartition time is a function of 
energy
density $\varepsilon$ only,
and therefore it is finite in the thermodynamic limit, although it
may diverge in the limit of vanishing energy density (see the
following).

Simulations for various energy densities $0.33 \le \varepsilon \le 
10^6$ 
allow us to study the variation of the equipartition time as a 
function of the
the energy density. In the low energy limit ($0.33 \le \varepsilon \le 
1.44$),
a good data collapse is obtained if the time
is rescaled by a typical time scale $\tau \sim \varepsilon^{-2}$ 
(see Fig. \ref{timeq} (a)). At high energy ($\varepsilon > 1000$)
the typical time scale is $\tau \sim \varepsilon^{-{1 \over 4}}$,
as it is clearly shown in Fig. \ref{timeq} (b). 
Let us also stress that $\varepsilon$ was also
found to be the relevant dependence of the 
largest Lyapunov exponent in the equipartition 
state~\cite{pettland,prl}.
Moreover, the maximal Lyapunov exponent shows two scaling laws at high
and low energy density that coincide with those here reported
for the equipartition time (namely, for its inverse)~\cite{CLP,prl}.
However, while for the Lyapunov the transition from one regime to 
the other is observed at $\varepsilon \simeq \beta^{-1}$ 
\cite{CLP,prl},
for the equipartition time the transition occurs at higher
energy : $\varepsilon \simeq 100$ for $\beta = 0.1$.
Power law divergences of the relaxation time have been recently 
reported
for several models of nonlinear oscillators and for different classes 
of 
initial conditions \cite{Licht95,PettCo96,Parisi}, although the large 
$N$ limit
has not been studied as carefully as in the present paper.

\section{RELATIONSHIP BETWEEN EXACT AND CHAOTIC BREATHERS}
\label{exact}

In the present section, we would like to pay attention to the
localized objects observed during the transient state, and study their
relationship with the already known localized solutions of nonlinear
lattices, that we call {\em exact} breathers.  We have seen that
the energy contained in a CB (and thus its frequency) can be very
high. For example, in chains of $N=512$ sites, we found breathers with
frequency $\omega \approx 3.5$ (when the maximal frequency associated
to the phonon band is 2). We would like to emphasize here that
the usually adopted Nonlinear Schr\"odinger
approximation~\cite{berman,maledetirusi} is not appropriate to
describe these objects.

The unstable $\pi$-mode gives rise spontaneously to a localized
breather-like excitation, qualitatively very similar to the exact
breathers obtained for the FPU chain\cite{newton}. The most evident
difference is that the former move in an erratic way in the lattice
and have a finite lifetime, while the latter are exactly periodic and
mainly static. Therefore it is natural to try to determine to what 
extent 
it is possible to compare them.

Another important difference between the self-created excitation and
the exact breathers concerns their dynamical stability, which has been
investigated with the aid of a Lyapunov analysis. On one hand exact
breathers are linearly stable and remain perfectly unchanged during
very long simulations. On the other hand, we show here that CBs
are strongly chaotic.

In order to compute the Lyapunov exponents of the system, we have used
the standard algorithm of Benettin {\em et al}~\cite{callyap}. 
However, 
we are here interested in the Lyapunov exponents at {\it short times}
instead of the asymptotic ones. We evaluate the cumulative
average corresponding to such exponents as a function of time.  In
Fig. \ref{cohlyap},   the evolution of the localization
parameter $C_0(t)$ is reported (panel (a)) together with the 
corresponding
cumulative average for the first four Lyapunov exponents (panel (b))
for a typical evolution starting from an unstable $\pi$-mode.

It is clear from Fig.~\ref{cohlyap}(b) that the localized object,
which has spontaneously appeared, is chaotic because its presence
(indicated by the peak in $C_0(t)$) is associated with a positive
maximal Lyapunov exponent.  Moreover, another typical feature of the
CB is that the running average for the maximal Lyapunov exponent has
an higher value during the transient than in the equipartition state.
The evolution of this running average is quite similar to that of
$C_0(t)$: there is an initial growth followed by a decrease at later
times.  Naively, one would usually expect that an higher degree of
chaoticity should be related to a higher degree of energy
equipartition: the present simulations clearly shows the opposite.

One should stress that the running average for the Lyapunov exponents
relaxes very slowly to the asymptotic value, because its value is
affected by the estimation at earlier times. A more efficient method
to obtain the asymptotic Lyapunov exponents is therefore to restart
the running averages {\em after} the transition to equipartition state
(i.e. after the decrease of $C_0$). The dash-triple dotted line of
Fig.~\ref{cohlyap}(b) shows that the convergence is much faster if
this is done.

Another interesting result is the fact that a gap is present in the
distribution of the Lyapunov exponents during the transient: the first
Lyapunov is clearly above the others whereas, when the system reaches
the energy equipartition state, the gap disappears (as shown in
Fig.~\ref{cohlyap}(b)) and the whole spectrum of Lyapunov exponents
is approximately linear~\cite{LPR}.

Furthermore, a measure of the localization of the Lyapunov vectors
shows that during the transient the first Lyapunov vector is localized
contrary to all others. This localization of the Lyapunov vector
disappears after that the equipartition is reached, since the energy 
density is below the strong stochasticity threshold, see~\cite{prl}.  
An interesting question is
to see whether this peculiar structure of the tangent space (i.e. the
gap in the spectrum and the localization of the first Lyapunov vector)
can be related to the properties of exact breathers.

Let us briefly recall that exact discrete breathers are time-periodic
and spatially exponentially localized solutions of the equations of
motion~(\ref{sub}). Their frequency is  always higher than the 
frequency
of the top of the phonon band. Calculating such a solution is
equivalent to looking for a fixed point of the stroboscopic $T$ map 
\begin{equation}(\{u_n,\dot{u}_n\})
(0) \mapsto T(\{u_n,\dot{u}_n\}) = \{u_n,\dot{u}_n\}(t_b)
\label{eq:period},
\end{equation}
where $t_b$ is the period of the solution.  This can be done with the
aid of a Newton process, using the tangent map $\partial T$.
The latter relates linearly an initial perturbation $\{\epsilon_i,
\dot{\epsilon}_i\}(0)$ to its image
$\{\epsilon_i,\dot{\epsilon}_i\}(t_b)$ where the perturbations
$\epsilon_i$ evolve according to the $N$ linearized equations of
motion
\begin{equation}
\ddot{\epsilon}_n = 
\left[1+3\beta(u_{n+1}-u_n)^2\right](\epsilon_{n+1}
-\epsilon_n) - \left[1+3\beta(u_n-u_{n-1})^2\right](\epsilon_n
-\epsilon_{n-1}).
\end{equation}
Starting from a sufficiently good approximation, the Newton method
converges to a periodic solution, satisfying Eq.~(\ref{eq:period}) 
up to machine precision\cite{newton}. One can then investigate the
linear stability of the breather solution with a standard Floquet
analysis, i.e. computing the eigenvalues of the $2N\times 2N$ matrix
$\partial T$ (a periodic solution is linearly stable when all
eigenvalues lie on the unit circle of the complex plane).

The main results concerning exact breathers in the FPU chain can
be summarized as follows. They exist for every frequency above the
phonon band; moreover the spatially antisymmetric solutions (centered
between two particles, as the generic example plotted in
Fig.~\ref{breather}(a) and  sometimes called $P$-modes~\cite{Page}) 
are
linearly stable, while the symmetric (the $ST$-modes\cite{ST}) are 
unstable. The spectrum of the Floquet matrix, discussed in detail in
\cite{CAF}, consists of a ``continuum'' of spatially extended
eigenmodes (the linear phonons) and a discrete part with a spatially
symmetric and exponentially localized mode. Due to the time 
reversibility 
of the solution, the real axis is a symmetry axis of the spectrum (if
$\lambda$ is an eigenvalue of $\partial T$, then $\lambda^*$ is also
an eigenvalue).  Fig.~\ref{breather}(b) shows schematically half of
the spectrum (the other half is its complex conjugate) with the
continuum and the discrete localized mode. As the problem is solved
for a {\em finite} chain, the continuum corresponds in reality to
$(N-1)$ modes with a higher density of modes close to the end of the
band.

Such linear modes out of the phonon band and with opposite symmetry to
the original solution, have been observed and studied in the framework
of Klein-Gordon chains. They are called {\em pinning} or {\em
translation} modes \cite{flachwillis,CAT,CA}. Indeed, an excitation of the
periodic solution in the direction of this mode produces an
oscillation of the center of energy or can even lead to a propagation
of the solution along the chain without radiative decay.  We have
checked that these features are also present in the $\beta$-FPU model.

The problem now is to link these exact solutions with the self-created
excitations which have been observed in the simulations. The essential
difference is that the former are linearly stable while the latter
are strongly chaotic.  However the comparison of the Lyapunov spectrum
of the CB and the Floquet spectrum of the exact breather solution
shows strong similarities: they both present a ``continuum'' of
extended states and an isolated spatially localized mode. A quite
natural hypothesis is that non-zero perturbations of the exact
breather could lead to a strong destabilization of the translation
mode. If this is true, one can understand why this instability is not
destructive for the breather: a perturbation along this direction only
leads to a coherent displacement of the breather and in addition, it
explains why, in the simulations, the non-linear excitation is very
seldom static.

In order to verify that the most unstable direction corresponds to the
translation mode of the exact solution, we have performed the
following test.  We study the evolution of an exact solution, once it
has been perturbed with a spatially antisymmetric gaussian noise all
over the lattice. The antisymmetry of the initial condition allows the
breather to remain at rest centered between two sites and this simplifies
the comparison.  The Lyapunov analysis reveals that the perturbed
state is now chaotic. The maximal Lyapunov eigenvector associated with
the perturbed solution and the Floquet vector corresponding to the
translational mode for the unperturbed case are shown in
Fig.~\ref{pert}. The two vectors are both symmetric and in perfect
agreement.  This is a convincing evidence
that the structure of the tangent space of a CB can be
qualitatively interpreted in terms of the Floquet spectrum of
``neighboring'' exact periodic solutions.

We would like to stress that the fact that breather-like excitations
move is not a sufficient condition for their chaoticity. Similarly to
what has already been found numerically for Klein-Gordon
chains~\cite{CA}, many exact mobile breather solutions can be
exhibited by the FPU model (see \cite{hori,FW,houle} or
\cite{flachwillis} for a discussion on
approximations of moving breathers). Such solutions satisfy the relation
\begin{equation} 
\{u_n,\dot{u}_n\}(T) = \{u_{n-1},\dot{u}_{n-1}\}(0)
\end{equation}
where $T$, the inverse of the velocity, is a multiple
of the period of internal vibration of the breather. One should also
note that a small amplitude phonon tail dresses these solutions, such
that one can consider that the breather is in equilibrium with the
emitted and the absorbed radiation (for more details see
Ref.~\cite{CA}).  Many of these solutions are linearly stable and were
checked to have zero maximal Lyapunov exponent; the chaotic regime is
reached only if the whole system is sufficiently perturbed.  The
direction of the translation mode is easily excitable, but is not the
direction responsible for chaoticity: thus mobile breathers are not
necessarily chaotic and strong chaos occurs only when other modes are
sufficiently excited.

\section{Formation and destruction of breathers}
\label{Formationdestruction}

At this stage, we have understood the connection of the CB with exact
FPU breathers but we should explain how the destabilization of the
$\pi$-mode can lead to only one localized solution. Indeed, as
attested by Fig.~\ref{greyscales}(a), the modulational instability
gives rise first to a few packets of energy but then, because of their
interactions, energy is concentrated in only one.

\subsection{The localization process}

In order to study this effect in a controlled manner, we have put four
exact breathers on the lattice: three moving ones with small amplitude
and a frequency $\omega=2.12$, that is just above the phonons' upper
band edge ($\omega=2$), and one at rest with a larger amplitude and a
frequency $\omega=2.75$.  As shown in Fig.~\ref{merging}, due to
collisions, the biggest breather successively absorbs the smaller ones
and gives rise to a moving large amplitude breather-like excitation.
Such a result is a generic example of the collision process in the FPU
chain and it explains why a single localized CB emerges during the
transient.

Let us emphasize that even if some localization processes have already
been reported in homogeneous nonlinear lattices~\cite{prllocali},
the process presented here is particularly interesting because of the
absence of the self-regulation process. 
Indeed, in the Klein-Gordon
systems studied in Ref.~\cite{prllocali,daumont}, the process
is regulated by a stronger pinning effect.
Discreteness provides a path to localization but is also 
responsible for the pinning effect. This 
is stronger for larger excitations (big breathers are easier trapped), 
and this does not allow a collapse of all the breathers into a 
single very large excitation.
For the FPU model, pinning effects are perhaps too small to be 
detected. Moreover, we have observed that the velocity of the CB slightly
increases with its energy. As a consequence, we see in
Fig.~\ref{greyscales} the emergence of only one large amplitude
breather-like excitation from the initial $\pi$-mode.

Is this effect present also in the thermodynamic limit ?  As the
lifetime $\tau$ and the velocity $v$ of a CB are finite, one could
predict that, for very long chains, the CB will probably not have
enough time to collect all the energy present in the system.  If, in
small chains, the breather could easily move through the system many
times, in very long chains this would not be the case.  One can
expect that above a critical chain length $L_c=v\tau$ more than one
CB will be observed in the transient preceding equipartition. To check
this, we have investigated the case with an initial amplitude $a=0.8$
for the $\pi$-mode; in this situation the typical velocity of the
final breather is $v\sim 0.2$ and its lifetime is $\tau\sim
10^4$. From these data the critical length $L_c$ is $\sim 2\cdot
10^3$.  For chains $N>10^3$ and $a=0.8$, additional simulations have
shown that this is indeed true: the energy is no longer localized in 
only one huge CB but in few of them.

A long chain could therefore be
considered as a juxtaposition of almost non-interacting sub-chains of
length $L_c$. And in each sub-chain, one single CB is created by
the modulational instability. This interpretation is consistent with the
observed saturation of the localization parameter $C_0$ when the size
of the system increases (Fig.~\ref{coherence}).  Indeed, a chain can
be considered as made up of 2 independent parts $A$ and $B$, and one
has $C_0(A+B)=(C_0(A)+C_0(B))/2 = C_0(A)$ ($C_0$ is an intensive-like
quantity). As discussed in Sec.~\ref{transition}, the lifetime is a 
function of
the energy density $\varepsilon$. Thus the critical length of
course depends on $\varepsilon$.

One can define two different regimes associated to the growth of the
excitation. In a first stage the main breather present in
the chain absorbs the smaller ones by collisions.  After this initial
stage the growth slows down.  This effect could be explained by an
observation made by Bang and Peyrard for the Klein-Gordon
equation~\cite{BangPeyrard}.  They noticed that the energy transfer
between localized modes depends on their relative energy difference:
the transfer is less efficient when the energy difference (or
frequency difference) increases.  For our system we have observed
that, after the first collisions, the lattice contains one big
breather with a rather high frequency and a large population of high
frequency phonons waves.  The presence of the latter is due to the
destabilization of the original $\pi$-mode via modulational
instability~\cite{prl}.  Correspondingly, the energy transfer-rate
between the phonons and the CB is still positive but reduced with
respect to the initial one.  We want to stress that this second regime
was detected because of the very small initial perturbations of the
$\pi$-mode, since for bigger perturbation this regime can easily be
missed~\cite{maledetirusi}.  In addition, in this paper we have
concentrated our attention on the energy region just above the
critical energy defined by Eq.~(\ref{ec}), where the FPU chain evolves
towards equipartition on a very long time scale.

\subsection{Breather's metastability}

Once the localization process is finished, a very large amplitude
breather is moving in the system and all the high frequency waves have
been absorbed. The hungry excitation cannot grow any more and this
regime corresponds to the plateau where $C_0$ has reached its
maximum. At later times, the excitation disappears and now we want to
explain how this phenomenon can take place.  We have seen that the
effect of collisions of the CB with high frequency modes leads to
absorption of the latter.  During such a scattering process, small
quantities of energy radiate towards low frequency modes.  Therefore,
after the initial stage the only phonons left in the system are the
low frequency ones. It is natural to suspect that the destruction of
the CB is related to their interaction with the low frequency modes.

In order to understand this interaction, we have performed numerical
experiments where one exact breather initially at rest collides with
the following wave packets of low frequency phonons centered on the
site $n_0$:

\begin{equation}
u_n(t)=A\ \cos(qn-\omega t)\ e^{-{(n-n_0)^2\over 2\ell_0^2}} \quad .
\end{equation}

We have taken $q=0.2$, $\ell=13$ and an amplitude
$A\in[0.5;3]$. Fig.~\ref{destruction} shows that the collision process
is very efficient in destroying the CB excitation.

As shown in Fig.~\ref{destruction}(a), the decrease of the energy of
the breather $E_B$ can be considered as linear at the earlier times
with a good approximation.  Moreover, a more detailed study of the
phenomenon shows that the amplitude of the wave packet determines the
destruction rate.  In particular, the slope of the decrease of $E_B$
is an exponential function of the amplitude. This result suggests that
a very complicated nonlinear mechanism is at the origin of the
interaction.  To give a more quantitative explanation we should study
the scattering on one phonon with the CB as done by Cretegny et
al\cite{CAF} for exact breathers at rest in the Klein-Gordon model.
However, here,  the study will be much more
complicated because small perturbations will 
easily put the FPU-breather in motion.

A possible explanation of the CBs metastability is the following: at 
the
beginning of the simulation, when only the $\pi$-mode is excited, and
even during the modulational instability process, the low frequency
phonons are not present.  During the growing process of the CB
the low frequency phonon band is populated due to
radiative processes. When all the high frequency phonons have been
absorbed by the CB, the excitation can only loose energy due to
the destructive interaction with the low frequency phonons. Therefore,
we can say that the CB ``die of starvation''. The destruction of the
CB is associated with a significant increase in the population of 
low
frequency linear waves. This transition corresponds to the final
decrease of the slope $S(t)$ plotted on Fig.~\ref{slope} and 
to the final relaxation to the equipartition state.

\section{CONCLUSION}
\label{conclusions}

We have seen that the evolution towards equipartition in the
$\beta$-FPU chain, starting from the $\pi$-mode as initial condition,
gives rise to a striking localization process.  The spontaneously
created excitations are moving breather-like excitations with a finite
lifetime and a chaotic dynamics.  The features of these transient
localized modes can be explained by exploiting 
the correspondence with exact breathers.  

We want also to stress that, contrary to the previous
belief, we have found that the localization mechanism originally
obtained in the Klein-Gordon case~\cite{prllocali,daumont}, is also a
transient state: the breathers obtained by collisions have
a very long lifetime but are indeed
CBs as characterized by the local Lyapunov exponent. The lifetime 
of the process is however much longer in this
case, because pinning effects on the excitation are stronger.

It is important to recall that the FPU equation was at the origin of
the rediscovery of the soliton~\cite{zabuskykruskal} in the continuum
limit, which justified the recurrence phenomenon observed by
Fermi-Pasta-Ulam ~\cite{FPU} (see also \cite{goedde2}).  
In the present paper, we have shown the
fundamental role played by a second family of excitations, the
breathers, in achieving the equipartition state on a discrete FPU
lattice.  The creation of a CB can be considered as an efficient
mechanism to transfer energy from high frequency modes to low
frequency phonons.

Let us also remark that the breathers are created from the high
frequency zone corresponding to zero group velocity.  This portion of
the phonon band was also recently found to be crucial for the
transition to equipartition \cite{shepeliansky}.  Indeed, the
transition to equipartition disappears if a renormalized FPU
Hamiltonian with a dispersion relation without zero group velocity
region is considered \cite{shepeliansky}.  As in electromagnetism and
string theory, the idea that this ingredient is
absolutely necessary to observe the transition to equipartition of
energy is another reason to justify a posteriori our study.

Moreover, we have found that the fundamental parameter for the
dynamics of the system is always the energy density $\varepsilon=E/N$
(i.e.  energy/volume). This is true not only for the value of the
maximal Lyapunov exponent (as already noticed in
Refs.~\cite{prl,pettland}) but also for the transient time towards
energy equipartition.  One should of course emphasize that this is the
only parameter which makes sense in the thermodynamic limit.
Therefore, the transition to equipartition happens in a finite time in
the thermodynamic limit, but this time diverges as an inverse power of
the energy density in the zero energy density (zero temperature)
limit.

As a final remark, this study emphasizes that the concept of a breather
is not only important for the energy localization in lattices, but is
also crucial to address one of the main classical problem of
statistical mechanics: the transition to equipartition.

\acknowledgements

 T.C. is supported 
by ``le Conseil de la R\'egion Rh\^one-Alpes'' with a grant 
``Emergence''. 
T.D. gratefully acknowledges EC for financial support with grant
ERBFMBICT961063.  This work is also part of the European contract
ERBCHRXCT940460 on ``Stability and universality in classical
mechanics''.  S.R. acknowledges the hospitality of the {\it Centro
Internacional de Ciencias} in Cuernavaca, M\`exico and the financial
support of the same Center and of CONACYT. 
A.T. thanks the
University of Wuppertal (Germany) and the Ecole Normale Sup\'erieure
of Lyon (France) for the nice hospitality. We would like to thank S. Aubry,
R. Livi and M. Peyrard for a careful reading of this manuscript. Part
of CPU time has been nicely supplied by the Institute of Scientific
Interchanges (ISI) of Torino.

\begin{figure}
\caption{Evolution of the local energy $E_n$ along the chain.
In Fig.(a), the horizontal axis indicates the position
along the chain and the vertical axis corresponds to
time (time is going upward). The grey scale goes from
$E_n=0$ (white) to the maximum $E_n$-value (black). The lower
rectangle corresponds to $0<t<3000$ and the upper one to
$5.994\ 10^5<t<6.10^5$. Figs. (b), (c) and (d) show the
instantaneous $E_n$ along the $N=128$ chain at three
different times. Note the difference in vertical amplitude.
The initial $\pi$-mode amplitude was $a=0.4$ (while
$a_c = 0.0317$).}
\label{greyscales}
\end{figure}

\begin{figure}
\caption{Evolution of $C_0(t)$ for chains
of various lengths: $N=32$,
128,  512, 1024, 2048 and  8192. Each curve corresponds to the average
over 20 simulations with the same energy density
$\varepsilon=E/N=1.44$ but with different random initial
noise added to the velocities. For all the reported simulations
$E > E_c$ {\protect\cite{alter}}.}
\label{coherence}
\end{figure}

\begin{figure}
\caption{Fig. (a) (resp. (b))  presents the evolution of the
slope $S(t)$ (resp. $\eta(t)$) versus time for chains
of various lengths;  $N=$32, 64, 128, 256 and 512. 
Each curve
corresponds to the average over 20 different initial conditions 
with the same energy density $\varepsilon=1.44$.}
\label{slope}
\end{figure}

\begin{figure}
\caption{In Fig. (a) (resp. (b)) the evolution of $\eta(t)$
versus the rescaled time $t\left({E\over N}\right)^2$ 
(resp. $t\left({E\over N}\right)^{1/4}$) is reported
for different energy densities. Each curve in Fig. (a) 
(resp. (b)) corresponds to the average over 20 
(resp. 50) different initial conditions for a chain 
with $N=512$ sites. The dot-dashed line reported in Fig. b
indicates the asymptotic value $\bar \eta \simeq 0.655$, which is  
theoretically estimated in Appendix B.}
\label{timeq}
\end{figure}

\begin{figure}
\caption{Fig. (a) presents the evolution of $C_0(t)$  and
Fig. (b) presents the cumulative average of the 
first four Lyapunov exponents when
the initial condition is a $\pi$-mode with $a=0.4$ in a
$N=128$ chain. 
The solid line corresponds to the first
Lyapunov  exponent (the dash-triple dotted line corresponds
to the value of the largest Lyapunov exponent after
the transient), the dotted one to the second, the dashed
one to the third and the dash-dotted line to the fourth. 
$t_2$ and $t_3$ in (a) are defined in Fig.~1.}
\label{cohlyap}
\end{figure}

\begin{figure}
\caption{Fig. (a) presents the spatial shape  of an exact
linearly stable FPU breather, solution of Eq.~(1) with a
frequency $\omega=2.5$. 
Fig. (b) shows the schematic repartition of its Floquet
eigenvalues on the complex plane.}
\label{breather}
\end{figure}

\begin{figure}
\caption{Shape of the translation mode. The solid line
corresponds to the eigenmode of the Floquet analysis whereas the stars
correspond to the first Lyapunov eigenvector. We have reported 
only the part of the vectors relative to the positions
when the internal phase of the breather is zero
(i.e. when the kinetic energy is minimal).}
\label{pert}
\end{figure}

\begin{figure}
\caption{Breathers' merging. Fig. (a) shows the energy evolution of
an exact breather initially at rest ($\omega=2.75$)
after collisions with three small exact moving breathers 
($\omega=2.12$).
Fig. (b) (resp. (c)) shows the
energy repartition at $t=0$ (resp. at $t = 2\cdot 10^4$).}
\label{merging}
\end{figure}

\begin{figure}
\caption{Breathers' destruction. Fig. (a) shows the evolution of
the energy of an initial exact static breather ($\omega=2.4$)
after collisions with a wave packet of phonons ($A=1.5,\ \ell_0=13,\ 
q=0.2$).
Fig. (b) (resp. (c)) shows the initial (resp. final) repartition of 
energy.}
\label{destruction}
\end{figure}

\appendix
\section{Calculation of the equipartition value $\bar C_0$
of the localization parameter}
\label{appen}
Introducing the usual parameter $\beta=1/k_BT$,
we obtain directly from Eq.~(\ref{clocal}), the asymptotic 
expression corresponding to energy equipartition
\begin{eqnarray}
\bar C_0=N{\displaystyle \sum_iE_i^2\over \left[\displaystyle 
\sum_i E_i\right]^2}
={\langle E_i^2\rangle\over \langle E_i\rangle^2}~,
\end{eqnarray}
where the spatial averages are 
\begin{eqnarray}
\langle E_i\rangle=
\langle {p_i^2\over 2}+{V_i\over 2}+{V_{i+1}\over
2}\rangle={1\over 2\beta}+{\langle V_i\rangle}~,
\end{eqnarray}
and
\begin{eqnarray}
\langle E_i^2\rangle&&=
\langle \left({p_i^2\over 2}+{V_i\over 2}+{V_{i+1}\over
2}\right)^2\rangle
={3\over 4\beta^2}+{1\over \beta}\langle V_i\rangle+
{\langle V_i^2\rangle\over 2}
+{\langle V_{i}\rangle^2\over 2}\quad.
\end{eqnarray}

For long chains, we can evaluate 
$\langle E_i\rangle$ and
$\langle E_i^2\rangle$ using the canonical ensemble.
Introducing the configurational partition function
\begin{equation}
F(\beta)=\int_{-\infty}^{+\infty}\ \exp\left[-\beta
V(x)\right]\ dx~,
\end{equation}
we have
 $\textstyle \langle V_i\rangle=-{1\over F}{\partial F\over
\partial\beta}$
and $\textstyle\langle V_i^2\rangle={1\over F}{\partial^2 F\over
\partial\beta^2}$.

One obtains therefore:
\begin{eqnarray}
\langle E_i\rangle
&&={1\over 2\beta}-{1\over F}{\partial F\over
\partial\beta}~,
\end{eqnarray}
and
\begin{eqnarray}
\langle E_i^2\rangle
&&={3\over 4\beta^2}
-{1\over \beta F}{\partial F\over
\partial\beta} +
{1\over 2F^2}\left({\partial F\over \partial\beta}\right)^2
+{1\over 2F}{\partial^2 F\over
\partial\beta^2}~.
\end{eqnarray}

In the pure harmonic case this gives
$\langle E_i\rangle={1 \over \beta}$ and $\langle E_i^2\rangle={7\over 
4\beta^2}$,
whereas in the pure quartic case we obtain
$\langle E_i \rangle={3 \over {4 \beta}}$ and $\langle E_i^2\rangle=
{19 \over {16\beta^2}}$.

In conclusion,
$\bar C_0={7\over 4}$ in the pure harmonic case and
$\bar C_0={19\over 9}\simeq 2.11$ in the pure quartic case.
But one can even compute the result using the complete
FPU-potential.
$V(x)={x^2\over 2}+\delta{x^4\over 4}$.

Using ${\cal K}_{1\over 4}$ the modified Bessel
function of the second kind, we have
\begin{eqnarray}
F(\beta)&&=\sqrt{1\over 2\delta}\ e^{\displaystyle
\beta\over\displaystyle  8\delta}\ {\cal
K}_{1\over 4}\left({\beta\over 8\delta}\right).
\end{eqnarray}
We obtain finally for
$\delta=0.1$,  the value 
\begin{equation}
\bar C_0\simeq1.795~,
\end{equation}
in excellent agreement with the numerical value.

\section{Estimation of the equipartition value $\bar \eta$
in the harmonic approximation}
\label{appen1}

We would like to give a theoretical estimation of the
indicator $\eta(t)$ (see Eq. (\ref{eta})) in the limit
$t \to \infty$, i.e. in the equipartition state
(the asymptotic value is indicated as $\displaystyle\lim_{t \to 
\infty} 
\eta(t) = \bar \eta$).

Let us rewrite the spectral entropy as
\begin{equation}
S = - \sum_{q=1}^{N/2}  \ p_q(t)\ \ln p_q(t)
\end{equation}
where $p_q(t)=\displaystyle{\phi_q\over\displaystyle \sum_{q}\phi_q}$.
Introducing
the energy per mode $e=\displaystyle{2\over N}\displaystyle
\sum_{q=1}^{N/2} \phi_q$, in the
equipartition state, we can easily derive
the following expression 
\begin{equation}
S = - {{ < \phi \ln \phi > } \over {e}} +\ln e + \ln \left({N\over 
2}\right)
\label{SS}
\end{equation}
The average appearing in (\ref{SS}) is 
$< \phi \ln \phi > =  \displaystyle{2 \over N}\displaystyle \sum_q 
\phi_q \ln \phi_q $,
and due to the equipartition the index $q$ has been neglected.

The corresponding expression for the quantity (\ref{eta}) is
\begin{equation}
\bar \eta = e  \quad {\rm e}^{- {1 \over e} < \phi \ln \phi >}
\end{equation}
Assuming that the fluctuations can be neglected, we obtain
\begin{equation}
<\phi \ln \phi> = <\phi> <\ln \phi> =<\phi> \ln <\phi>  = e \ln e
\label{elne}
\end{equation}
and $\bar \eta$ would be exactly one. However, the numerical value 
of $\bar \eta$ estimated in the low energy limit is quite 
different from one (namely, it is $\simeq $ 0.795). This indicates 
that
the inclusion of the fluctuations is fundamental to give a 
realistic estimate of $\bar \eta$ (this was firstly noticed in
\cite{goedde}). In order to consider the fluctuations, let us write
\begin{equation}
\ln \phi = \lim_{n \to 0} {{ \phi^n -1} \over n}~.
\end{equation}
This allows one to reexpress Eq. (\ref{elne}) as a function
of average of powers of $\phi$
\begin{equation}
<\phi \ln \phi> =\lim_{n \to 0} {{ <\phi^{n+1}> - e} \over n}
\end{equation}

The estimation of $\bar \eta$ is now reduced to the estimation
of terms of the type $< \phi^n >$, which can be again evaluated 
within the canonical ensemble, after a mode inverse temperature
$\beta^*$ is introduced. In the harmonic approximation
\begin{equation}
< \phi^{n} > = < \left( {|V_q|^2 + \omega^2 |X_q|^2} \over 2 
\right)^{n} >
= \sum_{k=0}^{n}  {{ n !} \over {k ! (n-k) !}} J_k J_{n-k} J_0^{-2}~,
\label{complic}
\end{equation}
with
\begin{equation}
J_k = \int_{-\infty}^{+\infty} dx \left( { x^2 \over 2} \right)^k 
{\rm e}^{-\beta^* {x^2 \over 2}} = \Gamma \left(k +{1 \over 2} \right) 
{ {\sqrt{2}} \over {(\beta^*)^{k+1/2}}}~,
\end{equation}
where $\Gamma$ is the Euler Gamma function. In particular, $e = < \phi 
> 
= (\beta^*)^{-1}$ and the expression (\ref{complic}) reduces to
\begin{equation}
< \phi^{n} > = e^n \Gamma(n+1) = e^n n !~.
\end{equation}
We are now able to give an expression for the spectral entropy.
\begin{equation}
<\phi \ln \phi> =\lim_{n \to 0} {{ e^{n+1} \Gamma(n+2) - e} \over n}
= e [\log(e)+\Gamma^\prime(2)]~,
\end{equation}
being $\Gamma^\prime(2) = 1 - \gamma$, where $\gamma \simeq 0.5772$
is the Euler constant. Finally we obtain
\begin{equation}
\bar \eta = {\rm e}^{\gamma -1} \simeq 0.655~.
\end{equation}
This results have been also obtained with a different approach
in \cite{goedde}. We notice that this value is in perfect
agreement with the numerical one found in the high energy
limit (see Fig. 4(b)), while it is an underestimation of 
the one found in the low energy limit (see Fig. 4(a)).

The origin of such a discrepancy is related to the structure
of the Hamiltonian when expressed in terms of normal modes.
Due to the quartic term in the potential, the mode interaction
matrix is no longer diagonal \cite{PR}. However, in the high energy 
limit 
a sort of "random phase approximation" should be valid for the terms 
appearing in the mode interaction matrix
and the off-diagonal terms should average to zero.
Therefore we are left, in the action-angle representation 
$(I_q,\theta_q)$, with diagonal terms of the form
\begin{equation}
H_{qq} = (\omega(q) I_q)+{{\beta^*} \over {2N}}
\omega(q)^2 I_q^2 < \cos^4 \theta_q >~.
\end{equation}
In the limit $N \to \infty$ the nonlinear term in $H_{qq}$
becomes negligible and the harmonic approximation is recovered.

In the low energy limit, the action-angle variables are
strongly correlated and a "random phase approximation"
is no longer valid. Therefore the off-diagonal terms
cannot be neglected and give a contribution (although
comparatively small) to $\bar \eta$. 


\begin{references}
\bibitem{alter}
$E_c(N)$ vanishes as $N^{-1}$, but $E_c(32)\simeq 1.04$ is not 
negligible for 
$\varepsilon=1.44$.
\bibitem{coreceta}
The asymptotic value $\bar \eta$, for low (resp. high) energy 
densities, is rather $\simeq 0.795$ (resp. $\simeq 0.655$) 
than 1 because of the fluctuations around the value $p_q=2/N$.
$\bar \eta$ shows a continuous transition from 0.795 to
0.655 for increasing $\varepsilon$. These discrepancies were already 
noticed in \protect\cite{goedde}. In Appendix B a more detailed
analysis of this point is reported.
\end{references}
\end{document}